\documentclass[twocolumn]{aastex631}
\usepackage{amssymb,amsmath}
\usepackage{graphicx}
\usepackage{xcolor,natbib}
\usepackage{multirow}
\usepackage[T1]{fontenc}
\usepackage{ae,aecompl}
\usepackage{newtxtext, newtxmath}
\usepackage{float}
\usepackage{subfigure}
\usepackage{longtable}
\usepackage{enumitem}
\usepackage{longtable,listings}
\usepackage{longtable}
\usepackage{booktabs}
\usepackage{subfigure}
\usepackage{multirow}
\usepackage[flushleft]{threeparttable}
\usepackage{parcolumns}
\def\oiii{[O~{\sc iii}]}

\submitjournal{ApJ}

\shorttitle{long-term optical variability}
\shortauthors{Zheng et al.}

\begin{document}
	
\title{On type 1 active galactic nuclei with double-peaked [O~{\sc iii}]. II. properties of long-term optical variability}

\correspondingauthor{Xueguang Zhang; Qirong Yuan}
\email{xgzhang@gxu.edu.cn; yuanqirong@njnu.edu.cn}

\author{Qi Zheng}
\affiliation{School of Physics and Technology, Nanjing Normal University, No. 1,	Wenyuan Road, Nanjing, 210023, P. R. China}

\author{Xingyv Zhu}
\affiliation{School of Physics and Astronomy, Beijing Normal University, Beijing, 100875, P. R. China}

\author{Xueguang Zhang$^{*}$}
\affiliation{Guangxi Key Laboratory for Relativistic Astrophysics, School of Physical Science and Technology,
	GuangXi University, No. 100, Daxue Road, Nanning, 530004, P. R. China}

\author{Qirong Yuan$^{*}$}
\affiliation{School of Physics and Technology, Nanjing Normal University, No. 1,	Wenyuan Road, Nanjing, 210023, P. R. China}

\begin{abstract}
Double-peaked \oiii~profiles could potentially indicate kiloparsec-scale dual AGNs. 
We analyze long-term optical light curves of 35 type 1 AGNs with such features from our recent catalog in Zheng et al. (2025).
These light curves are obtained from the Catalina Sky Survey and modeled using a Damped Random Walk (DRW) process.
A control sample of 210 normal type 1 AGNs matched in redshift, intrinsic luminosity, and black hole mass is also studied.
If the double-peaked \oiii~are caused by two type 1 AGNs (dual type 1 AGN), then the combined variability from the two AGNs would be expected to differ from that of a single type 1 AGN.
However, there is no statistically significant difference in the variability timescale $\tau$ and intrinsic variability amplitude $\sigma$ between these double-peaked AGNs and the control sample of 210 normal type 1 AGNs. Crucially, computer simulations reveal that dual AGN systems systematically produce lower variability amplitudes than single AGNs, which is inconsistent with the observed variability properties of our double-peaked \oiii~sample. 
Moreover, simulations suggest that the fraction of dual type 1 AGNs is $\sim$ 3\%, indicating that double-peaked \oiii~may not be a reliable indicator of dual type 1 AGNs in these systems. However, this does not rule out the possibility that some objects may still host dual AGNs involving other combinations, such as type 1+type 2 AGNs.
Future studies with larger samples and higher-quality light curves will help clarify the true nature of these systems.
\end{abstract}


\section{Introduction}
Dual active galactic nuclei (AGNs) have attracted significant attention over the years as key systems for understanding black hole accretion and AGN activity during the process of galaxy mergers \citep{Ko03,Ge07,De19,Pe23,Li23,Sa24}. 
\citet{Zh04} identified an object exhibiting double-peaked \oiii~emission lines along with two radio cores at 8.4 GHz, and proposed that double-peaked narrow emission lines could serve as a promising method for detecting dual AGN systems.
In the dual AGN scenario, the two sets of narrow emission lines
arise from the orbital motion of two AGNs with their own narrow line regions within a single galaxy formed through a merger \citep{Ba13}.
Then numerous studies have focused on objects exhibiting double-peaked narrow emission lines, especially the [O~{\sc iii}]$\lambda\lambda$4959,5007, identified as potential candidates for kpc-scale dual AGNs \citep{Xu09,Wa09,Mc11,Ge12,Se21,Zh24}.
Through the combination of imaging and spectroscopy, several AGNs exhibiting double-peaked narrow emission lines have been confirmed as dual AGN systems \citep{Liu10,Fu11,Ro11,Sh11,Li13,Co15}.

Double-peaked narrow emission lines could also be produced by narrow line region kinematics, such as
biconical outflow \citep{Wh04} or disk rotation of the narrow line region \citep{Sm12}, related to a single AGN.
Several studies \citep{Ml15,Ne16,Ru19} proposed that outflows are the dominant factor affecting double-peaked narrow emission lines.
\citet{Sh11} argued that the majority of double-peaked \oiii~in AGNs is related to narrow line region kinematics.

The presence of distinctive narrow emission-line profiles suggests that these AGNs may constitute a unique subclass within AGN population.
As a fundamental property of AGNs, optical variability offers important clues about the structure and dynamics of their emission regions \citep{To00,Mu11,Ki20,Lo23}.
Given that typical type 2 AGNs tend to show no variability in long-term light curves \citep{An93,Yi09,Ne15,Sa17,Lo23}, our analysis primarily focuses on type 1 AGNs.
If a double-peaked \oiii~profile indeed originates from two type 1 AGNs, their combined light curve may exhibit the superposition of two independent variability signals. As a result, the variability properties of these systems may differ statistically from those of normal single-nucleus type 1 AGNs.
By analyzing the variability features of type 1 AGNs with double-peaked \oiii~and normal type 1 AGNs, we aim to evaluate whether the double-peaked objects show evidence of dual activity. This approach can allow us to assess the likelihood of dual type 1 AGNs in these systems using relatively accessible photometric monitoring data, without requiring expensive high-resolution follow-up observations.

Among the different techniques applied to describe AGN variability, \citet{Ko10} introduced a refined mathematic technique for estimating the parameters of the Damped Random Walk (DRW) process, and demonstrated its effectiveness in describing AGN variability.
Modeling AGN variability as DRW process could significantly enhance the understanding of the mechanisms driving AGN luminosity variability in ultraviolet/optical bands \citep{Ma10}.
\citet{Zu11,Zu13} developed JAVELIN, a publicly available code, which applies the methodology proposed by \citet{Ko10} to model AGN variability using the DRW process.
\citet{Ke09} introduced a groundbreaking approach to studying AGN variability by employing stochastic diffusion processes, specifically continuous autoregressive (CAR) process, to directly analyze the inherently non-periodic nature of AGN light curves.
The basic parameters of the intrinsic variability amplitude $\sigma$ and the intrinsic variability timescale $\tau$, determined by CAR/DRW process, could be applied to describe AGN intrinsic variability.
Then, more complex stochastic process models, such as the continuous-time autoregressive moving-average processes \citep{Ke14,Ka17,St22}, were provided, which allowed for a wider variety of power spectral density shapes and offer improved fitting performance, provided the light curve data is of sufficiently high quality.

Thus far, there is no comprehensive statistical analysis of
the long-term variability characteristics of type 1 AGNs with double-peaked narrow emission lines.
A recent study by \citet{Zh25} identified 62 low-redshift type 1 AGNs exhibiting double-peaked \oiii~profiles based on Data Release 16 of the Sloan Digital Sky Survey (SDSS). The majority of these objects also have available light curves from the Catalina Sky Survey (CSS) \citep{Dr09}.
The CSS provides photometric data for 500 million objects with V magnitudes ranging from 11.5 to 21.5, covering an area of 33,000 square degrees.
Large samples of type 1 AGNs from SDSS were presented by \citet{sh112} and \citet{Ra20}, offering a valuable control sample for comparison with type 1 AGNs exhibiting double-peaked \oiii~profiles. This provides an opportunity to perform a statistical analysis of their long-term optical variability and explore potential differences in the properties of emission regions between double-peaked and normal type 1 AGNs.
While higher-order Continuous-time AutoRegressive Moving Average (CARMA) models \citep{Ke14,Yu22} may capture AGN variability more accurately, they involve more free parameters that are difficult to constrain, especially with limited data quality. This complexity also poses challenges for the subsequent simulation, as poorly constrained priors may lead to unphysical results.
Here, we apply the DRW process by the JAVELIN code from \citet{Zu11,Zu13} to model the large statistical samples of type 1 AGNs.
This method has been widely applied in numerous AGN variability studies \citep[e.g.,][]{Ki19, Gu22, Ma24}, demonstrating its reliability and effectiveness for characterizing optical light curves.
The DRW process could effectively capture AGN optical variability across all timescales and provide parameters useful for predicting diverse AGN properties and their emission region characteristics \citep{Zu13}.

In this manuscript, the main results of the type 1 AGNs with double-peaked \oiii~and normal type 1 AGNs are shown in Section 2, the necessary simulation and corresponding results  are presented in Section 3, and the summary and conclusions are given in Section 4.
We have adopted the cosmological parameters of $H_{0}=70 {\rm km/s/Mpc}$, $\Omega_{\Lambda}=0.7$ and $\Omega_{\rm m}=0.3$.
\section{Main Results}
The sample of 62 type 1 AGNs with double-peaked \oiii, reported in our recent work \citep{Zh25}, is applied as our main parent sample to study variability of AGNs with double-peaked \oiii.
This sample is drawn from a large database with strict selection criteria for type 1 AGNs with double-peaked profiles. The type 1 AGNs from both \citet{Sm10} and \citet{Ge12} that meet our selection criteria have already been incorporated into our current sample. 
Then, light curves of the 62 type 1 AGNs with double-peaked \oiii~in our sample are searched in the Catalina Sky Survey \citep{Dr09}, with position radius within  0.002 degrees (7.2$^{\prime\prime}$). 
The long-term variability of the 62 objects, as observed in the CSS data, can be described using the publicly available JAVELIN code \citet{Zu11,Zu13}.
To characterize the long-term variability of the 62 objects observed in the CSS data, we apply the JAVELIN code \citep{Zu11,Zu13}, which implements the DRW process commonly used to characterize AGN optical variability.
This analysis employs the DRW process, characterized by two fundamental parameters: the intrinsic variability amplitude $\sigma$ and the intrinsic variability timescale $\tau$.

The JAVELIN code uses Markov Chain Monte Carlo (MCMC) sampling with logarithmic priors over a wide range of parameter values (0<$\tau$<$10^5$ days, 0<$\sigma$<$10^2$ mag day$^{-0.5}$; \citealt{Zu11,Zu13}), which are sufficient to encompass the typical variability timescales and amplitudes observed in AGNs.
This approach can ensure reliable determination of the posterior distributions, and provide well-constrained final parameters with associated confidence intervals.
In final, 35 objects in our sample are collected
with determined DRW parameters (ln$\sigma$ and ln$\tau$) exceeding their uncertainties by at least a factor of 3. 
The light curves and the corresponding best fitting results by the DRW process of the 35 type 1 AGNs with double-peaked \oiii~are shown in Figure \ref{fig1}.
Meanwhile, the derived DRW parameters of the 35 type 1 AGNs in the our sample are presented in Table \ref{tab1}. 
After correcting for redshift, the mean values of ln$\tau$ and ln$\sigma$ for the 35 objects are 5.34$\pm$0.89 and –2.41$\pm$0.59, respectively. The corresponding uncertainties of ln$\tau$ and ln$\sigma$ here and hereafter are standard deviation. The corresponding median values of ln$\tau$ and ln$\sigma$ are 5.43 and –2.50, respectively.

\begin{figure*} 
\centering\includegraphics[width=18cm,height=21cm]{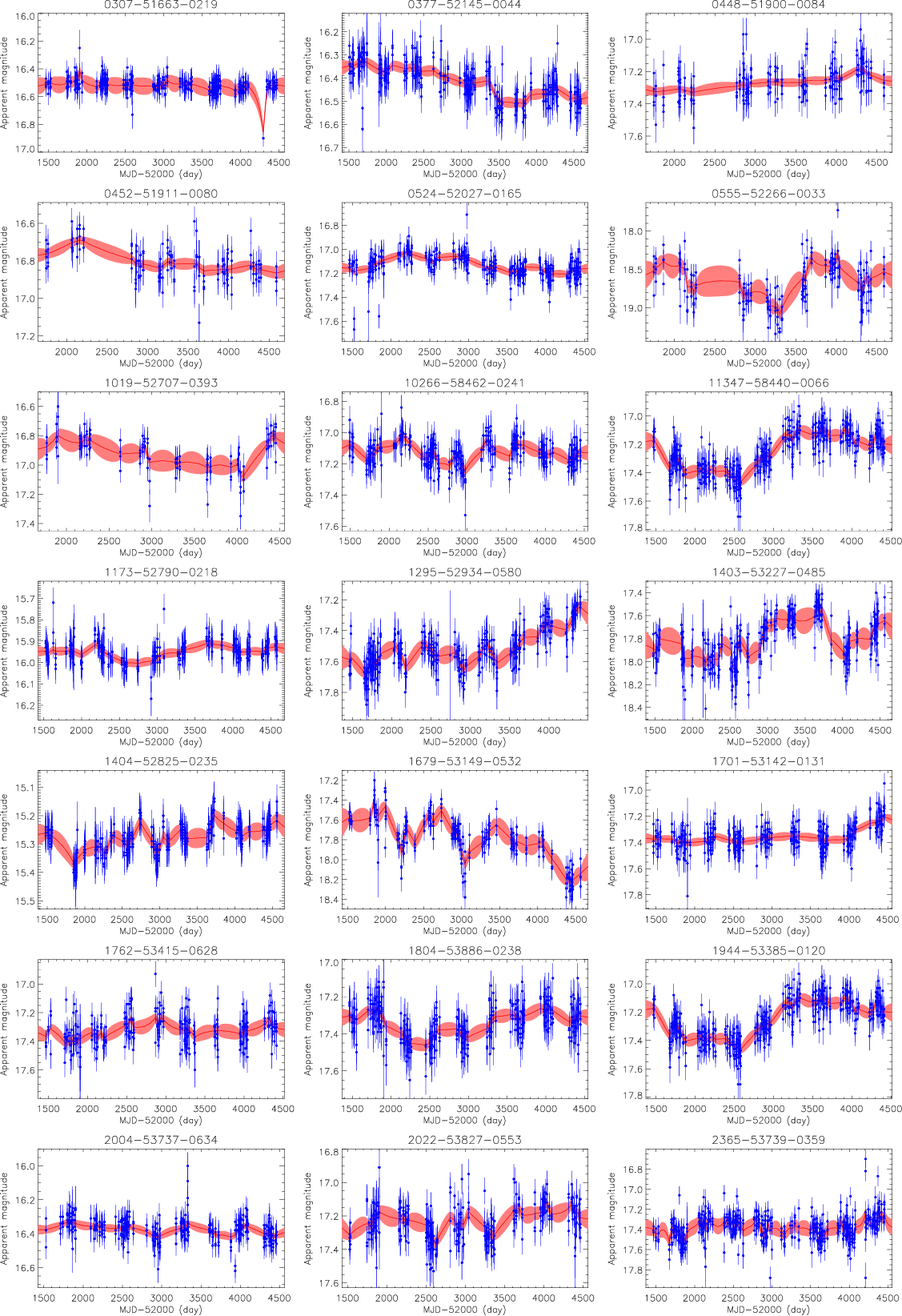}
\caption{The best descriptions to the long-term CSS (solid blue dots with error bars) light curves of the 35 type 1 AGNs with double-peaked \oiii~in our sample determined by the JAVELIN code.
In each panel, Plate-Mjd-Fiberid is shown in the title, the solid red line represents the best fitting results to the CSS light curve, with corresponding shaded regions indicating the 1$\sigma$ confidence intervals.			
}
\end{figure*} 
\setcounter{figure}{0}
\begin{figure*} \centering\includegraphics[width = 18cm,height=15cm]{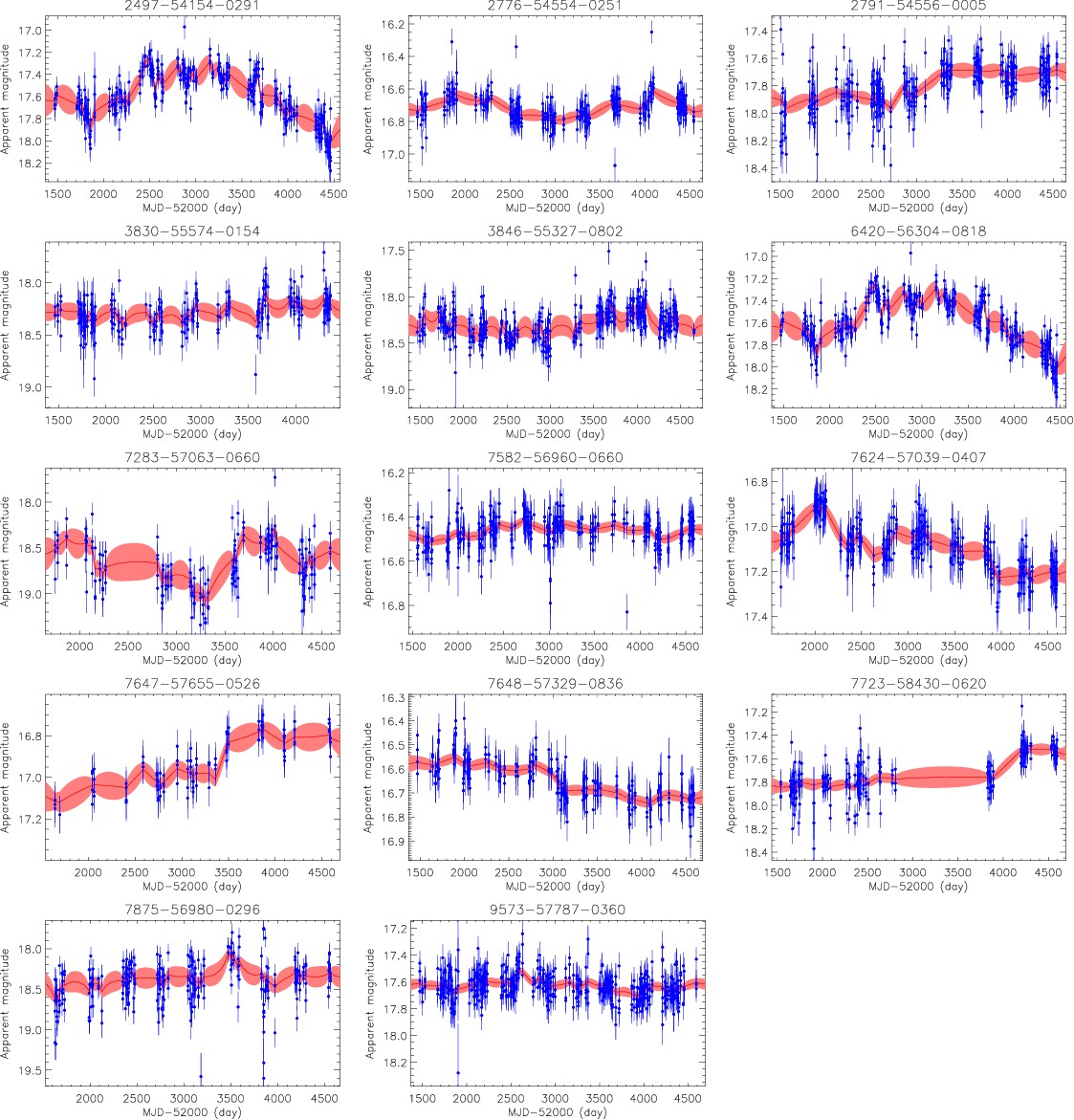} \caption{-- to be continued}
	\label{fig1} \end{figure*} \setcounter{figure}{1}

In addition to analyzing the DRW-determined variability properties of the 35 double-peaked \oiii, it is necessary to determine long-term variability properties for a large sample of normal type 1 AGNs, in order to find clearer variability difference between the type 1 AGNs with double-peaked \oiii~ and the normal type 1 AGNs.
As discussed in \citet{Ke09}, the variability of type 1 AGNs is strongly correlate with luminosity and black hole mass. Besides, redshift indicates the evolutionary history of central black hole, so the effect of different redshift is also considered here.
The intrinsic luminosity of the 35 type 1 AGNs exhibiting double-peaked \oiii~is derived from the spectroscopic results presented in \citet{Zh25}. The black hole masses are estimated using the equation of broad H$\beta$ emission lines (from Table 7 of \citet{Zh25}) in \citet{Ve06}, with interstellar extinction corrections by \citet{Fi99} after assuming an intrinsic Balmer decrement of 3.1. The information of redshift, black hole mass and intrinsic luminosity of the 35 type 1 AGNs are listed in Table \ref{tab1}.

\begin{figure*}
	\centering\includegraphics[width=18cm,height=6cm]{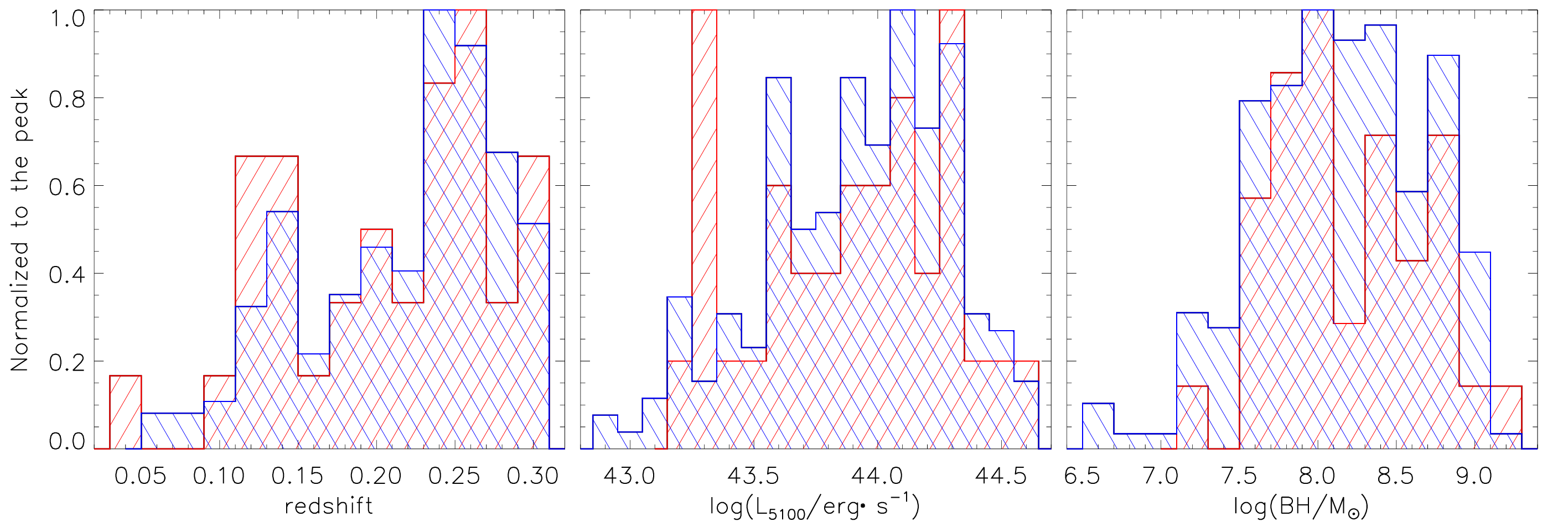}
	\caption{The distributions of redshift, intrinsic luminosity and black hole mass for the 35 type 1 AGNs with double-peaked \oiii~(in red color) and the 210 normal type 1 AGNs(in blue color).}
	\label{fig3}
\end{figure*}

The control sample of 210 normal type 1 AGNs here is collected from the QSO catalog \citep{sh112,Ra20}, as shown in Figure \ref{fig3}, the distributions of redshift, intrinsic luminosity and black hole mass of the normal type 1 AGNs sample is similar to the 35 type 1 AGNs with double-peaked \oiii.
To ensure a close match between the control sample and the double-peaked sample in terms of redshift, intrinsic luminosity, and black hole mass, we select, for each object in the double-peaked sample, the six nearest neighbors in the three-dimensional parameter space defined by these parameters.
This selection strategy effectively aligns the one-dimensional distributions of redshift, intrinsic luminosity, and black hole mass between the two samples.
We further verify this similarity by performing independent one-dimensional Kolmogorov–Smirnov (K-S) test \citep{Ko33,Sm48} on each parameter, resulting statistically similar with the probability of 93.71\%, 95.46\%, and 93.27\%, respectively.

\begin{figure*}
\centering\includegraphics[width=8cm,height=5cm]{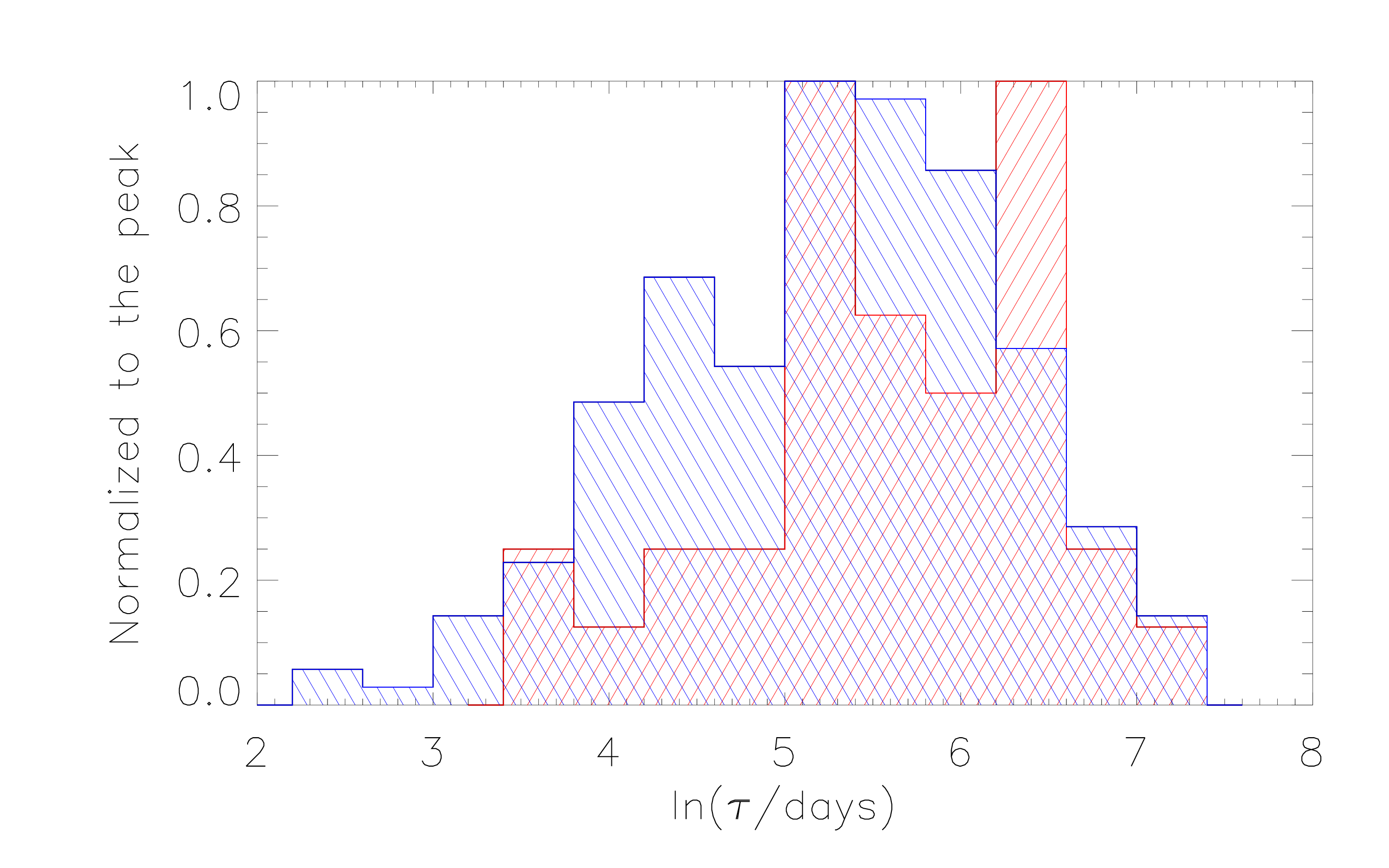}
\centering\includegraphics[width=8cm,height=5cm]{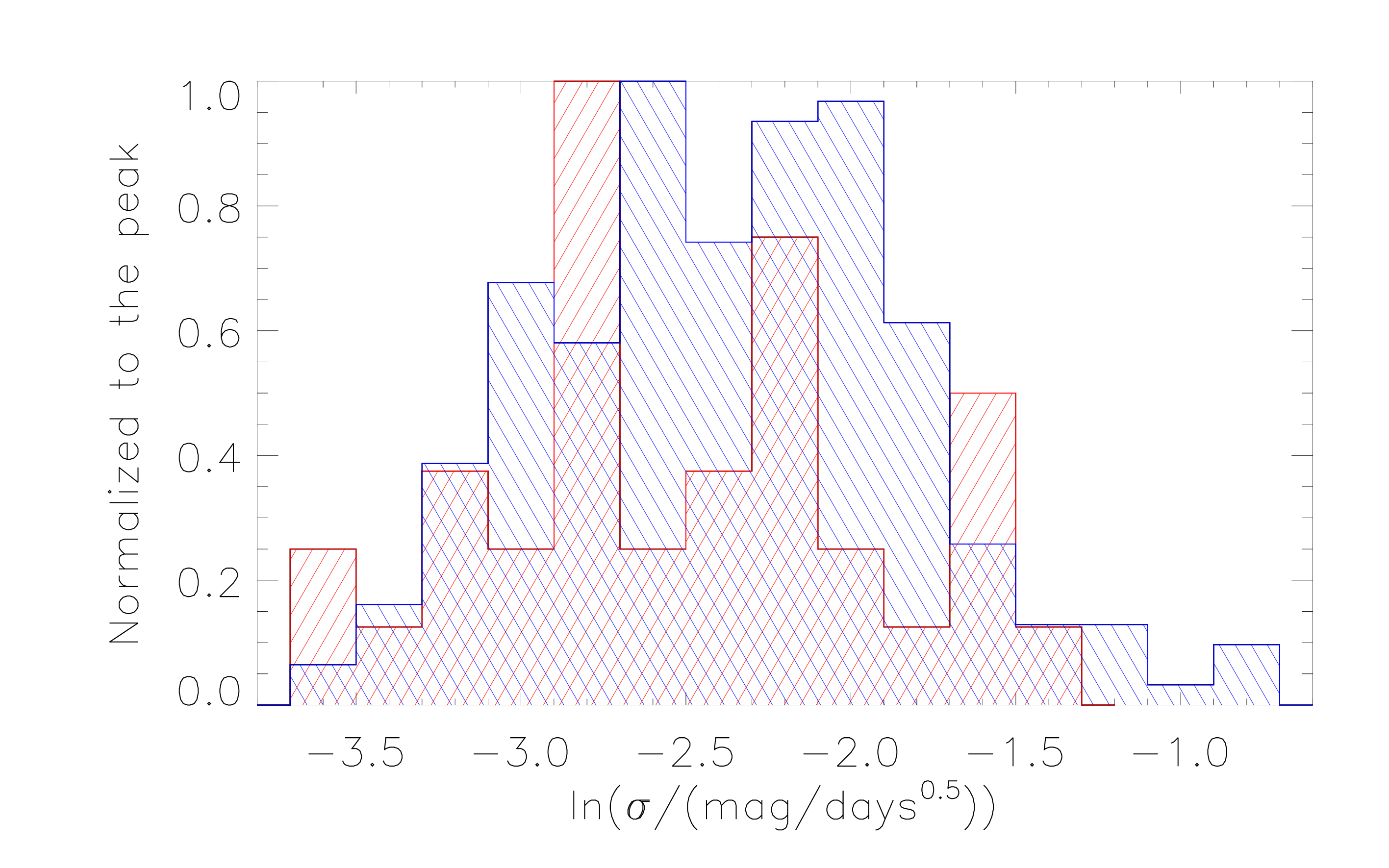}
\caption{The distributions of ln$\tau$ (left panel) and ln$\sigma$ (right panel) in the rest-frame for the 35 type 1 AGNs with double-peaked \oiii~(in red color) and the 210 normal type 1 AGNs (in blue color).}
\label{fig4}
\end{figure*}

The light curves of the 210 normal type 1 AGNs are collected from the CSS, and modeled as described above.
All 210 normal type 1 AGNs in the sample exhibit obvious long-term variability, with determined DRW parameters (ln$\sigma$ and ln$\tau$) exceeding their respective uncertainties by at least a factor of 3.
The mean ln$\tau$ and ln$\sigma$ values (after correcting for redshift) of the 210 objects are 5.05$\pm$0.99 and -2.27$\pm$0.56, respectively.
The distributions of ln$\tau$ and ln$\sigma$ of the two samples are shown in Figure \ref{fig4}.
To assess whether the DRW process parameters differ significantly between the 35 type 1 AGNs with double-peaked \oiii~and the control sample of 210 normal type 1 AGNs, we first perform the Student's t-test \citep{St08}, which assumes normal distributions and equal variances, to compare their mean values. The results show no statistically significant difference in the mean ln$\tau$ and ln$\sigma$ values, with p-values of 9.81\% and 18.09\%, respectively. To complement this, we also apply the Mann–Whitney U test \citep{Ma47}, which does not assume normal distribution or equal variances, to compare the median values between the two samples. 
The median ln$\tau$ and ln$\sigma$ values (after correcting for redshift) of the control sample are 5.15 and –2.22, respectively.
The Mann–Whitney U test similarly shows no significant difference in the median values of ln$\tau$ and ln$\sigma$ between the two samples, with p-values of 9.79\% and 22.65\%, respectively.

Furthermore, we systematically vary the matching sequence of our control sample and confirm that the mean (median) values of ln$\tau$ and ln$\sigma$ show no significant differences between the two samples. This consistency across different matching orders demonstrates that our results are not attributable to selection bias in the matching procedure.

\begin{table*}[htbp]
\centering
\caption{Basic parameters of the 35 type 1 AGNs with double-peaked \oiii}
\label{tab1}
\begin{tabular}{cccccccccccc}  
\hline
\textbf{Plate-Mjd-Fiberid} & \textbf{ln$\sigma$} & \textbf{ln$\tau$}&\textbf{z} & \textbf{$ L_{5100\AA}$} & \textbf{$ M_{\rm BH}$} &\textbf{Plate-Mjd-Fiberid} & \textbf{ln$\sigma$} & \textbf{ln$\tau$}&\textbf{z} & \textbf{$ L_{5100\AA}$} & \textbf{$ M_{\rm BH}$}
\\ \hline
0307-51663-0219	&	-2.8$\pm$0.2	&	4.3$\pm$0.5 	&	0.29 	&	44.09 	&	7.82 	&	2004-53737-0634	&	-3.5$\pm$0.3 	&	5.0$\pm$0.7 	&	0.13 	&	43.25 	&	7.61 	\\
0377-52145-0044	&	-2.8$\pm$0.3	&	6.7$\pm$0.7 	&	0.13 	&	44.04 	&	7.92 	&	2022-53827-0553	&	-2.7$\pm$0.2 	&	5.0$\pm$0.6 	&	0.21 	&	43.78 	&	8.70 	\\
0448-51900-0084	&	-3.2$\pm$0.3	&	6.0$\pm$0.9 	&	0.26 	&	44.11 	&	7.64 	&	2365-53739-0359	&	-2.6$\pm$0.1 	&	3.5$\pm$0.5 	&	0.11 	&	43.30 	&	7.92 	\\
0452-51911-0080	&	-2.9$\pm$0.3	&	6.2$\pm$0.9 	&	0.16 	&	43.33 	&	8.39 	&	2497-54154-0291	&	-1.6 $\pm$0.2 	&	5.7$\pm$0.5 	&	0.15 	&	43.31 	&	7.92 	\\
0524-52027-0165	&	-2.8$\pm$0.3	&	6.0$\pm$0.8 	&	0.23 	&	44.33 	&	7.59 	&	2776-54554-0251	&	-2.9 $\pm$0.3 	&	5.5$\pm$0.7 	&	0.11 	&	43.35 	&	7.74 	\\
0555-52266-0033	&	-1.6$\pm$0.2	&	5.0$\pm$0.6 	&	0.29 	&	44.10 	&	8.42 	&	2791-54556-0005	&	-2.2$\pm$0.3 	&	6.3$\pm$0.8 	&	0.26 	&	44.01 	&	8.01 	\\
1019-52707-0393	&	-2.4$\pm$0.2	&	5.4$\pm$0.8 	&	0.14 	&	43.61 	&	7.93 	&	3830-55574-0154	&	-2.4$\pm$0.2 	&	4.1$\pm$0.7 	&	0.26 	&	43.89 	&	8.04 	\\
10266-58462-0241	&-2.9$\pm$0.2	&	5.0$\pm$0.5 	&	0.24 	&	44.27 	&	8.86 	&	3846-55327-0802	&	-2.2$\pm$0.1 	&	3.7$\pm$0.5 	&	0.29 	&	43.93 	&	7.80 	\\
11347-58440-0066	&	-2.1$\pm$0.3	&	6.5$\pm$0.6 	&	0.27 	&	44.30 	&	8.82 	&	6420-56304-0818	&	-1.6$\pm$0.3 	&	5.8$\pm$0.6 	&	0.15 	&	43.69 	&	8.16 	\\
1173-52790-0218	&	-3.5$\pm$0.3	&	5.5$\pm$0.8 	&	0.12 	&	43.81 	&	7.23 	&	7283-57063-0660	&	-1.6$\pm$0.2 	&	5.1$\pm$0.6 	&	0.29 	&	43.92 	&	8.36 	\\
1295-52934-0580	&	-2.1$\pm$0.3	&	6.3$\pm$0.7 	&	0.24 	&	44.20 	&	8.32 	&	7582-56960-0660	&	-3.5$\pm$0.2 	&	5.3 	$\pm$	0.7 	&	0.14 	&	43.58 	&	7.55 	\\
1403-53227-0485	&	-1.9$\pm$0.2	&	5.4$\pm$0.5 	&	0.23 	&	43.65 	&	8.84 	&	7624-57039-0407	&	-2.4$\pm$0.3 	&	6.4$\pm$0.6 	&	0.30 	&	44.57 	&	7.87 	\\
1404-52825-0235	&	-3.1$\pm$0.2 	&	5.2$\pm$0.6 	&	0.05 	&	43.19 	&	7.87 	&	7647-57655-0526	&	-2.1$\pm$0.3 	&	6.5$\pm$0.8 	&	0.25 	&	44.55 	&	8.30 	\\
1679-53149-0532	&	-1.5$\pm$0.3 	&	6.1$\pm$0.7 	&	0.21 	&	44.14 	&	9.16 	&	7648-57329-0836	&	-2.8$\pm$0.3 	&	7.0$\pm$0.6 	&	0.21 	&	44.16 	&	8.32 	\\
1701-53142-0131	&	-2.8$\pm$0.3 	&	6.3$\pm$0.8 	&	0.29 	&	44.33 	&	8.58 	&	7723-58430-0620	&	-2.2$\pm$0.3 	&	6.6$\pm$0.7 	&	0.18 	&	43.63 	&	9.03 	\\
1762-53415-0628	&	-2.9$\pm$0.2 	&	4.7$\pm$1.2 	&	0.19 	&	43.42 	&	7.99 	&	7875-56980-0296	&	-2.0$\pm$0.2 	&	4.3$\pm$0.9 	&	0.23 	&	43.51 	&	7.80 	\\
1804-53886-0238	&	-2.7$\pm$0.3 	&	5.9$\pm$0.6 	&	0.24 	&	44.30 	&	8.90 	&	9573-57787-0360	&	-3.2$\pm$0.2 	&	4.7$\pm$0.6 	&	0.25 	&	44.03 	&	8.60 	\\
1944-53385-0120	&	-2.2$\pm$0.3 	&	6.4$\pm$0.6 	&	0.27 	&	44.44 	&	8.83 	&		&				&				&		&		&		\\	
\hline
\end{tabular}
\begin{tablenotes}
\item
Notice: Columns 1 and 7: Plate, MJD, Fiberid of spectroscopic observation;
Columns 2 and 8: the derived DRW parameter for the observed-frame logarithmic variability amplitude ($\sigma$, in units of mag/day$^{0.5}$); Columns 3 and 9: the derived DRW parameter for the observed-frame logarithmic characteristic timescale ($\tau$, in units of days);  Columns 4 and 10: the redshift; Columns 5 and 11: the logarithmic luminosity at rest-frame 5100 \AA ($ L_{5100\AA}$, in unit of erg/s); Columns 6 and 12: the virial black hole mass log($\rm M_{\rm BH}$/$\rm M_{\odot}$).
\end{tablenotes}
\end{table*}

\section{Simulation}
Type 1 AGNs exhibiting double-peaked \oiii~represent a peculiar subclass that may originate from dual AGN systems. 
Besides, the observed double-peaked narrow emission lines could potentially originate from other physical mechanisms unrelated to dual AGNs, such as a rotating disk model \citep{Xu09,Sm12} or biconical outflows \citep{Ne16,Co18}. 
However, these alternative scenarios are unlikely to have a significant impact on long-term optical variability. 
Kinematic features such as disk rotation or bipolar outflows primarily reshape the emission line profiles through changes in the velocity field \citep{Liu10b,Wa19}. In contrast, long-term optical variability is generally attributed to processes in the accretion disk around the central black hole \citep{Ke09,Ma10,Ji20} and is not expected to be strongly influenced by these large-scale kinematic structures. This difference may provide a useful diagnostic for exploring the origin of double-peaked narrow emission lines in AGN spectra.
While the DRW and/or CAR processes are mathematically well-defined, our goal here is to simulate the realistic combination of two observed AGN light curves in flux space, accounting for irregular sampling and finite durations—factors difficult to model analytically. The variability properties of superimposed light curves depend on relative phases, amplitudes, and sampling, which introduces additional complexity. Therefore, we employ simulations to empirically assess their statistical behavior under conditions matching our observations.
This simulation is motivated by the need to understand whether the composite variability signatures from dual type 1 AGNs might differ from those of single AGNs.

\begin{table*}
\caption{The simulation results of the composite and original light curves}
\centering
\label{tab4}
\begin{center}
\begin{tabular}{ccccccccccccc}
\hline
sample & $\bar{\sigma_{c}}$ & $\bar{\sigma_{o}}$ &$p_{\bar{\sigma}}$  & $\widetilde{\sigma_c}$ & $\widetilde{\sigma_o}$ & $p_{\widetilde{\sigma}}$
& $\bar{\tau_{c}}$ & $\bar{\tau_o}$ &$p_{\bar{\tau}}$  & $\widetilde{\tau_c}$ & $\widetilde{\tau_o}$ & $p_{\widetilde{\tau}}$ \\ \hline
All &-1.73$\pm$0.45&-1.55$\pm$0.48& <$10^{-10}$
&-1.68&-1.48&<$10^{-10}$&6.03$\pm$0.56&6.05$\pm$0.57&0.64&6.05&6.08&0.19\\
group(a)&-1.61$\pm$0.43&-1.43$\pm$0.48&<$10^{-10}$&-1.55&-1.33
&<$10^{-10}$&5.74$\pm$0.51&5.76$\pm$0.53&0.22&5.74&5.75&0.25\\
group(b)&-1.61$\pm$0.42&-1.43$\pm$0.46&<$10^{-10}$&-1.55&-1.33&<$10^{-10}$&5.76$\pm$0.50&5.78$\pm$0.53&0.03&5.76&5.79&0.02\\
group(c) &-1.85$\pm$0.44&-1.68$\pm$0.47&<$10^{-10}$&-1.80&-1.60
&<$10^{-10}$ &6.35$\pm$0.48 &6.36$\pm$0.47 & 0.33& 6.37 & 6.38& 0.44\\
group(d) & -1.86$\pm$0.43 & -1.67$\pm$0.47 & <$10^{-10}$ & -1.81& -1.60 & <$10^{-10}$ & 6.33$\pm$0.46 & 6.34$\pm$0.47 & 0.41 & 6.36 & 6.37 & 0.24 \\  \hline
\end{tabular}
\begin{tablenotes}
\item[*] Column 1: the sample of simulation; Column 2 (Column 3): the mean ln$\sigma$ values of composite (original) light curves; Column 4: the p-value, determined by Student's t-test, of the same mean ln$\sigma$ values between the composite and original light curves; Column 5 (Column 6): the median ln$\sigma$ values of composite (original) light curves; Column 7: the p-value, determined by Mann–Whitney U test, of the same median ln$\sigma$ values between the composite and original light curves; Column 8 (Column 9): the mean ln$\tau$ values of composite (original) light curves; Column 10: the p-value, determined by Student's t-test, of the same mean ln$\tau$ values between the composite and original light curves; Column 11 (Column 12): the median ln$\tau$ values of composite (original) light curves; Column 13: the p-value, determined by Mann–Whitney U test, of the same median ln$\tau$ values between the composite and original light curves;
\end{tablenotes}
\end{center}	
\end{table*}

\begin{figure*}
	\centering\includegraphics[width=8cm,height=5cm]{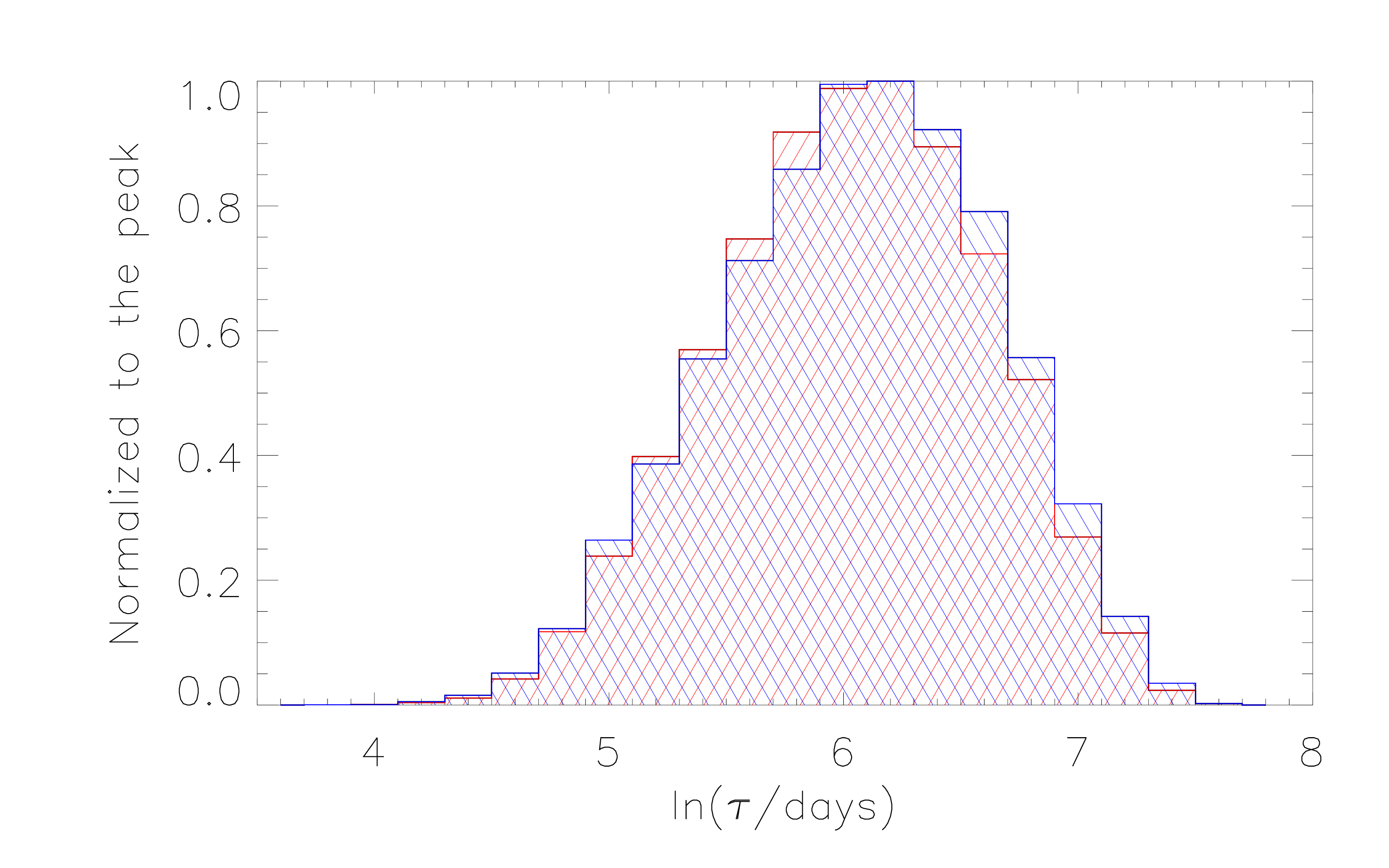}
	\centering\includegraphics[width=8cm,height=5cm]{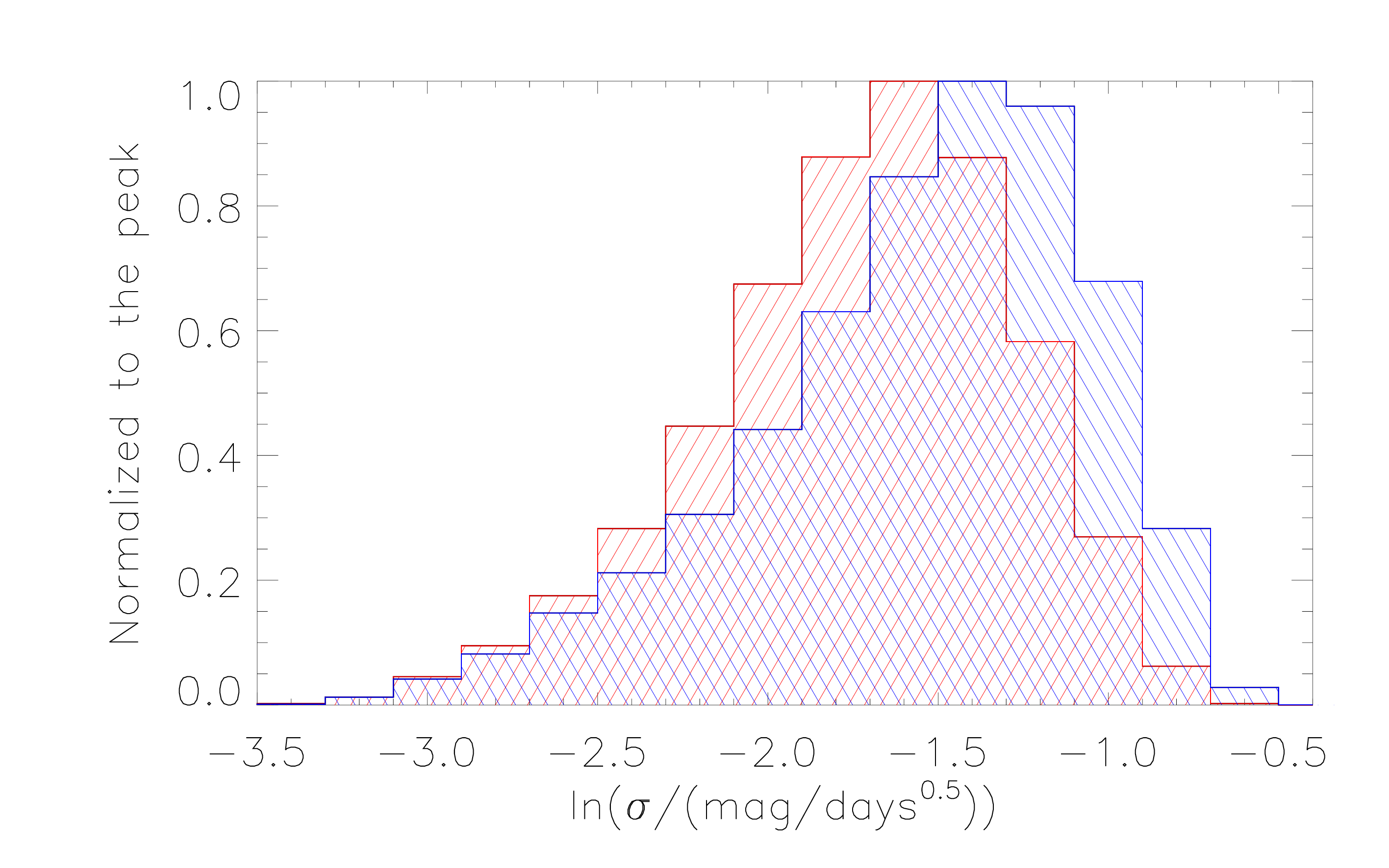}
	\caption{The distributions of ln$\tau$ (left panel) and ln$\sigma$ (right panel) for the 30,000 simulations.
		The histogram filled by blue lines represents the input parameters and the histogram filled by red lines represents the output parameters.}
	\label{fig5}
\end{figure*}

This simulation is structured as follows: (1) Light curve generation: generate two fake AGN light curves $X(t)$ using CAR process $dX(t)=-\frac{1}{\tau}X(t)+\sigma\sqrt{dt}\epsilon (t)+bdt, \tau ,\sigma, t >0$ ($\epsilon(t)$ as a white noise process, and $bdt$ as the mean value of $X(t)$) \citep{Ke09} with parameters sampled from: characteristic timescale $\tau \in [100,\ 1000]~\text{day}$ and variance $\in [0.004,\ 0.184]~\text{mag}^2$ ($\sigma$ constrained by  $\tau\sigma^2/2=$variance), and the time information $t$ here is randomly selected from observational data in the CSS. Then, random magnitude ($\in [14,\ 19]~\text{mag}$) is added to the two zero-mean light curves as their mean values.
(2) Single AGN components analysis: apply the JAVELIN code in each fake light curve, and reserve fake light curve with its intrinsic variability parameters (mentioned as input ln$\tau$ and ln$\sigma$ here and hereinafter) exceeding three times their respective uncertainties.
(3) Construction of composite light curve: combine the two fake light curves in flux space using the equation of $m_{\rm total}=-2.5\log_{10}(10^{-0.4m_{1}}+10^{-0.4m_{2}})$ ($m_{\rm total},~m_{1}$ and $m_{2}$ are the magnitudes of composite light curve and two individual fake light curves, respectively), and then analyze the composite light curve with JAVELIN code. The composite light curve with its intrinsic variability parameters (mentioned as output ln$\tau$ and ln$\sigma$ here and hereinafter) exceeding three times their respective uncertainties is retained.
(4) Statistical validation: repeat the aforementioned procedure for 30,000 times to ensure statistically robust results.
The simulation results are shown in Figure \ref{fig5} and the Table \ref{tab4}. Obviously, the mean (median) ln$\tau$ value of the original fake light curves is similar to the mean (median) ln$\tau$ of the composite light curves, while the mean (median) ln$\sigma$ value of the original fake light curves is significantly larger than the mean (median) ln$\sigma$ of the composite light curves.

\begin{figure*}
\centering\includegraphics[width=18cm,height=5.6cm]{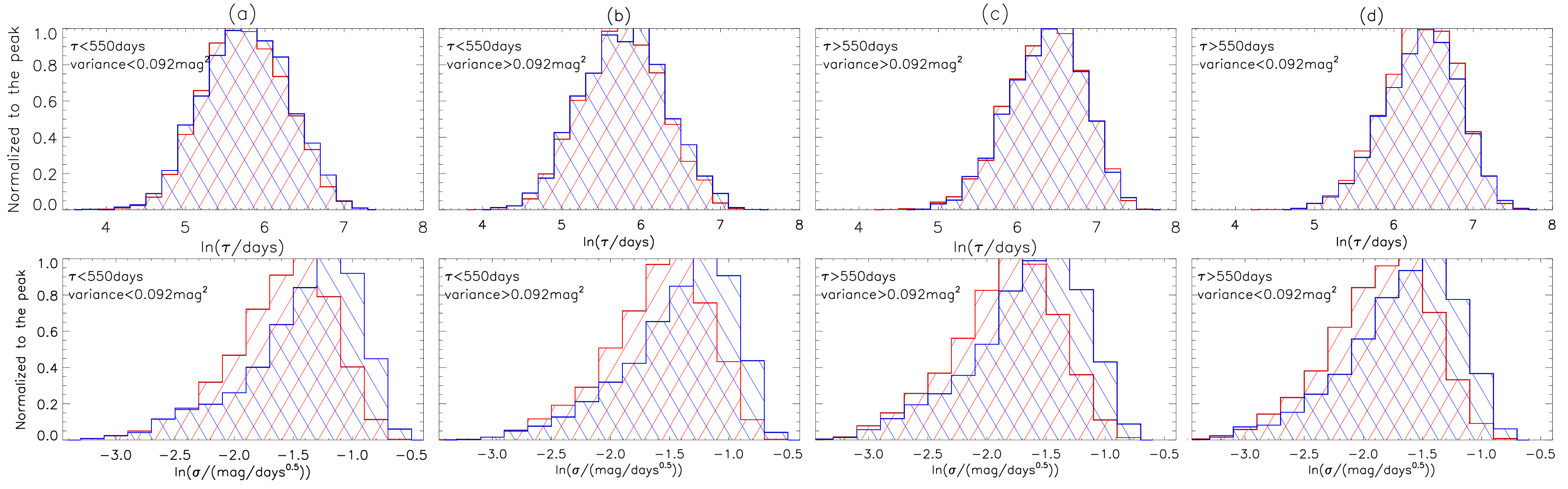}
\caption{The distributions of input parameters (in blue color) and output parameters (in red color). 
Top panels: the distributions of ln$\tau$; Bottom panels: the distributions of ln$\sigma$. The criteria of each subsample are listed in the top-left corner.
}
\label{fig6}
\end{figure*}

To systematically test whether the $\tau$ and $\sigma$ of the composite light curves are influenced by the input parameters, we conduct a stratified analysis of our 30,000 simulations based on two key input parameter thresholds: (1) variance > or < 0.092 mag$^2$ and (2) $\tau$ > or < 550 days.
According to the two groups of input parameters that meet the same condition above, this classification yield four distinct subgroups, which enable us to perform comparative analysis between the input and output ln$\tau$ (ln$\sigma$) distributions for each subgroup. 
The subdivision criteria are chosen to correspond to the median values of the respective parameter distributions, ensuring balanced sample sizes across subgroups while effectively probing parameter space regions where potential systematic effects might be most pronounced. 
The distributions of $\tau$ and $\sigma$ for the four subgroups (group(a), (b), (c) and (d)) are shown in Figure \ref{fig6}, and corresponding results are shown in Table \ref{tab4}.

In brief, after categorizing the simulation results into four distinct groups based on different conditions, we find that there is no significant difference in the mean and median ln$\tau$ values between the input and output ln$\tau$. In contrast, compared to the input ln$\sigma$ values of single-AGN components, there is a consistent and statistically significant reduction in output ln$\sigma$ values of composite light curves across all four groups.
It should be noticed that these results are based on the fundamental assumption that both AGNs in the dual AGN system are type 1 AGNs with observable optical variability. In reality, if double-peaked \oiii~originates from a dual AGN system comprising a type 1 AGN and a type 2 AGN (where the latter shows negligible optical variability due to obscuration), the resulting composite light curve would effectively preserve the characteristic variability of the type 1 component alone. 
Accounting for this mixed-type scenario, the dual AGN systems with at least one type 1 AGN would still exhibit lower variability amplitude $\sigma$ than single type 1 AGNs.

However, after accounting for the effects of redshift, intrinsic luminosity, and black hole mass, the 35 type 1 AGNs with double-peaked \oiii~in our sample show no large difference in variability timescale $\tau$ and variability
amplitude $\sigma$ compared to normal type 1 AGNs in \citet{sh112,Ra20} sample.
This observational result appears inconsistent with the simulation findings, where dual type 1 AGN systems consistently produce smaller characteristic $\sigma$ in composite light curves.

\begin{figure*}
	\centering\includegraphics[width=16cm,height=8cm]{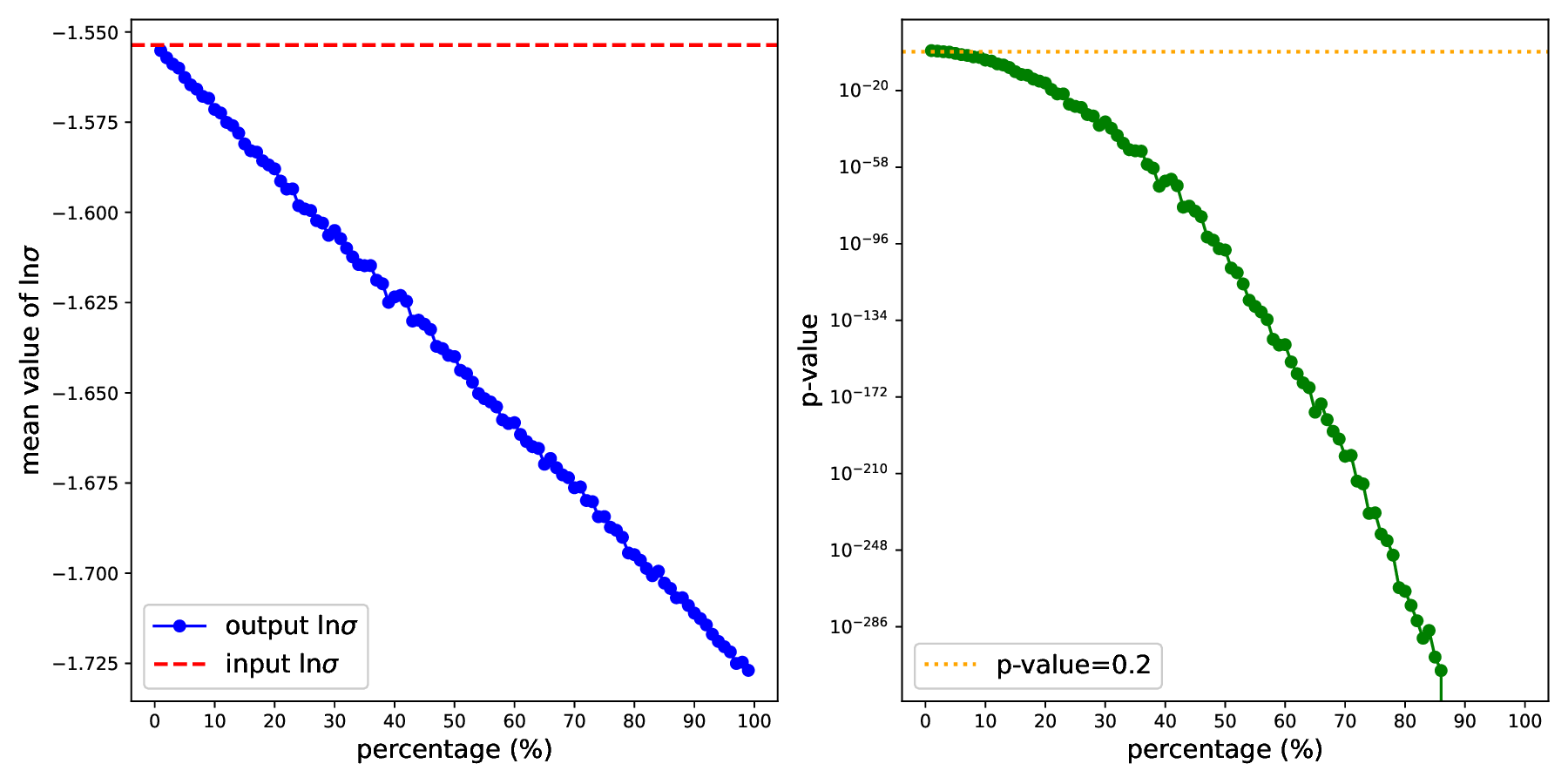}
	\caption{Left panel: the relation of proportion of dual type 1 AGN components and the mean ln$\sigma$ values (blue dots) determined by simulation results; Right panel: the Student's t-test results (green dots) between single-AGN components and single-AGN+dual type 1 AGN components. 
		The dashed red line and dashed orange line represent the mean value of input ln$\sigma$ and p-value=0.2, respectively.
	}
	\label{fig7}
\end{figure*}

To investigate the potential dual type 1 AGN fraction among the sample of 35 type 1 AGNs with double-peaked \oiii, we perform an analysis of the 30,000 simulations. The proportion of composite light curves is systematically varied while the fake light curves from single-AGN components are maintained as the remainder.
Then, the Student's t-test is applied to compare the mean variability amplitudes $\sigma$ between single-AGN components and single+dual type 1 AGN components, and the mean ln$\sigma$ values and the corresponding Student's t-test results are shown in Figure \ref{fig7}.
The results reveal that when 3\% of the light curves contain dual type 1 AGN components, the mean ln$\sigma$ values show no significant difference (p-value:0.20), indicating statistical indistinguishability from the pure single-AGN case.
And when the proportion is 5\%, the p-value will drop to 0.03.
It suggests that the actual dual type 1 AGN fraction in type 1 AGN with double-peaked \oiii~populations likely is about 3\%. 
Since the simulated results divided into four distinct groups show that the $\sigma$ values are significantly smaller in the composite light curves compared to the original ones, while $\tau$ values remain largely unchanged, only the $\sigma$ values are analyzed here; additionally, as the dual type 1 AGN proportions obtained from Student's t-test and Mann-Whitney U test are consistent, only the Student's t-test results are presented.

In conclusion, the apparent discrepancy suggests that the double-peaked \oiii~in these type 1 AGNs may not mainly originate from dual type 1 AGN systems. 
But this conclusion maybe currently limited by the small sample size. In the future, statistically complete samples of type 1 AGNs with double-peaked \oiii~ will enable more robust testing of this scenario through improved measurement precision.

\section{Summary and Conclusions}
The long-term light curves of 35 type 1 AGNs with double-peaked \oiii~from \citet{Zh25} are collected from the CSS, and then analyzed by the DRW process.
After considerations of necessary effects of redshift, intrinsic luminosity, and black hole mass, there is no large difference in variability timescale $\tau$ and the intrinsic variability
amplitude $\sigma$ between the 35 type 1 AGNs with double-peaked \oiii~and a control sample of 210 normal type 1 AGNs from \citet{sh112,Ra20} sample. 
However, the computer simulations demonstrate that dual type 1 AGNs systematically produce lower $\sigma$ compared to single AGN systems. This reduced variability amplitude is inconsistent with the observed properties of type 1 AGNs exhibiting double-peaked \oiii. 
The simulation results indicate that about 3\% of type 1 AGNs with double-peaked \oiii~are likely to represent dual type 1 AGN systems.
Therefore, based on our current sample, double-peaked \oiii~may constitute an unreliable diagnostic for identifying dual type 1 AGN systems.

This study highlights the potential of utilizing optical variability as a diagnostic tool to explore the physical origins of double-peaked narrow emission line profiles. 
Although our current analysis does not strongly support the dual type 1 AGN scenario, it does not rule out the possibility of dual AGNs involving other combinations, such as type 1 and type 2 AGNs. Future investigations based on larger samples of type 1 AGNs with double-peaked \oiii~profiles, combined with higher-quality light curves, will be crucial for drawing more definitive conclusions.

\section*{Acknowledgements}
We gratefully acknowledge the anonymous referee for giving us constructive comments and suggestions to greatly improve our paper.
This work is supported by the National Natural Science Foundation of China (Nos. 12273013, 12173020, 12373014), and the Postgraduate Research \& Practice Innovation Program of Jiangsu Province (Grant No.KYCX25\_1932). We have made use of the data from SDSS DR16.
The CSS web site is (\url{http://nesssi.cacr.caltech.edu/DataRelease/}). 
The web site of JAVELIN code is(\url{https://github.com/nye17/javelin/}).


\end{document}